\documentclass[editor]{raa}            

\usepackage{graphicx,times}             

\begin{document}

   \title{Characterising motion types of G-band bright points in the quiet Sun
   \footnotetext{$*$ Supported by the National Natural Science Foundation of China.}
}

   \volnopage{Vol.0 (2015) No.0, 000--000}      
   \setcounter{page}{1}          

   \author{Yun-Fei Yang
        \inst{1,2,3}
   \and Hui-Xue Qu
        \inst{1}
   \and Kai-Fan Ji
        \inst{1}
    \and Song Feng
        \inst{1,2,3}
   \and Hui Deng
        \inst{1}
   \and Jia-Ben Lin
        \inst{2}
   \and Feng Wang
        \inst{1,4}
   }

   \institute{Faculty of Information Engineering and Automation / Yunnan Key Laboratory of Computer Technology Application, Kunming University of Science and Technology,
   Kunming 650500, China; {\it yfyangkmust@gmail.com}\\
        \and
             Key Laboratory of Solar Activity, National Astronomical Observatories, Chinese Academy of Sciences, Beijing 100012, China\\
        \and
             Key Laboratory of Modern Astronomy and Astrophysics, Nanjing University, Ministry of Education, Nanjing 210093, China\\
         \and
         Yunnan Observatory, Chinese Academy of Sciences, Yunnan 650011, China\\
   }

   \date{Received 2014 April 30; accepted 2014 July 29}

\abstract{We study the motions of G-band bright points (GBPs) in the quiet Sun to obtain the characteristics of different motion types. A high resolution image sequence taken with the Hinode/Solar Optical Telescope (SOT) is used, and GBPs are automatically tracked by segmenting 3D evolutional structures in a space-time cube. After putting the GBPs that don't move during their lifetimes aside, the non-stationary GBPs are categorized into three types based on an index of motion type. Most GBPs that move in straight or nearly straight lines are categorized into a straight motion type, a few moving in rotary paths into a rotary motion, and the others fall into a motion type we called erratic. The mean horizontal velocity is 2.18$\pm$0.08\,$\rm km$ $\rm s^{-1}$, 1.63$\pm$0.09\,$\rm km$ $\rm s^{-1}$ and 1.33$\pm$0.07\,$\rm km$ $\rm s^{-1}$ for straight, erratic and rotary motion type, respectively. We find that a GBP drifts at a higher and constant velocity during its whole life if it moves in a straight line. However, it has a lower and variational velocity if it moves in a rotary path. The diffusive process is ballistic-, super- and sub-diffusion for straight, erratic and rotary motion type, respectively. The corresponding diffusion index ($\gamma$) and coefficients ($K$)  are 2.13$\pm$0.09 and 850$\pm$37\,$\rm km^{2}$ $\rm s^{-1}$, 1.82$\pm$0.07 and 331$\pm$24\,$\rm km^{2}$ $\rm s^{-1}$, 0.73$\pm$0.19 and 13$\pm$9\,$\rm km^{2}$ $\rm s^{-1}$. In terms of direction of motion, it is homogeneous and isotropical, and usually persists between neighbouring frames, no matter what motion type a GBP belongs to.
}

   \authorrunning{Y.-F. Yang, H.-X. Qu, K.-F. Ji, S. Feng, H. Deng, J.-B. Lin \& F. Wang }            
   \titlerunning{Characterising motion types of G-band bright points in the quiet Sun}  

   \maketitle

%
%
\section{Introduction}           
\label{sect:intro}

Photospheric bright points moving in intergranular dark lanes are clearly visible in G-band observations, and are often referred to as G-band bright points (hereafter GBPs). Recent studies show that a GBPs motion is determined mainly by the buffeting motion of granules. The convective motions of granules push and squeeze GBPs violently, resulting in chaotic motion of GBPs (Berger et al.~\cite{Berger98a}; van Ballegooijen et al.~\cite{van98}; M\"{o}stl et al.~\cite{Mostl06}; Giannattasio et al.~\cite{Giannattasio13}; Keys et al.~\cite{Keys14}). These small-scale motions are thought to play an important role in heating the upper solar atmosphere by twisting and braiding magnetic flux tubes, generating kinetic, Alfv\'{e}n waves and so on (Cranmer~\cite{Cranmer02}; Cranmer \& van Ballegooien~\cite{Cranmer05}; Klimchuk~\cite{Klimchuk06}; de Wijn et al.~\cite{de09}; Zhao et al.~\cite{Zhao09}; Balmaceda et al.~\cite{Balmaceda10}).

High resolution observations have led scientists to study the motions of GBPs. Results show mean horizontal velocities of 1-2\,$\rm km$ $\rm s^{-1}$ with maximum values of 7\,$\rm km$ $\rm s^{-1}$ (Muller et al.~\cite{Muller94}; Berger et al.~\cite{Berger98b}; M\"{o}stl et al.~\cite{Mostl06}; Utz et al.~\cite{Utz10}; Keys et al.~\cite{Keys11}). Isolated GBPs move fast at birth, then decrease to their lowest velocities in the middle stage of their lifetime, and then they accelerate again in the decay stage until their eventual disappearance (Yang et al.~\cite{Yang14}). The direction of motion usually persists from one time interval to the next (Nisenson et al.~\cite{Nisenson03}; Bodn\'{a}rov\'{a} et al.~\cite{Bodn13}).

The random motion of GBP is commonly described in terms of turbulence. One important component of turbulence is diffusion; where the dispersion mechanism is defined by the diffusion index ($\gamma$) and the efficiency of this mechanism is given by the diffusion coefficient ($K$). Earlier studies interpreted their results as indicative of normal- or sub-diffusive processes (Berger et al.~\cite{Berger98a}; Cadavid et al.~\cite{Cadavid99}). However, recent studies, both using G-band images and magnetograms from ground and space observations, agree on a super-diffusive dynamic regime with $\gamma$ of 1.2 $-$ 1.7 (Abramenko et al~\cite{Abramenko11}; Chitta et al.~\cite{Chitta12}; Giannattasio et al.~\cite{Giannattasio14}; Jafarzadeh et al.~\cite{Jafarzadeh14}; Keys et al.~\cite{Keys14}). The diffusion coefficient measures the rate of increase in the dispersal area in units of time. The values reported in previous literatures lie between 0.87\,$\rm km^{2}$ $\rm s^{-1}$ and 350\,$\rm km^{2}$ $\rm s^{-1}$ (Berger et al.~\cite{Berger98a}; Utz et al.~\cite{Utz10}; Manso Sainz et al.~\cite{Manso11}; Giannattasio et al.~\cite{Giannattasio14}; Jafarzadeh et al.~\cite{Jafarzadeh14}).

Another component of turbulence is vortex motion. Vortices occurring in simulations have been suggested as a primary candidate mechanism, with roots in the photosphere, for energy transport from the solar interior to the outer layers of the solar atmosphere (V\"{o}gler et al.~\cite{Vogler05}; Carlsson et al.~\cite{Carlsson10}; Fedun et al.~\cite{Fedun11}; Moll et al.~\cite{Moll11}; Shelyag et al.~\cite{Shelyag11}; Shelyag et al.~\cite{Shelyag13}). Shelyag et al. (\cite{Shelyag11}) simulated G-band images and showed a direct connection between magnetic vortices and rotary motions of photospheric bright points, and suggested that there may be a connection between the magnetic bright point rotation and small-scale swirl motions observed higher in the atmosphere. Vortex-type motions also have been measured by tracking bright points in high-resolution observations of the photosphere. Bonet et al. (\cite{Bonet08}) traced the motions of bright points which follow spiral paths on the way to being engulfed by a downdraught. Goode et al. (\cite{Goode10}) noted that colliding granules create a vortex into which the encircled bright points enter and spin around each other. Additionally, Wedemeyer-B\"{o}hm et al. (\cite{Wedemeyer12}) demonstrated, using a series of co-spatial images at different atmospheric layers, that ``magnetic tornadoes" in the chromosphere and transition region results in rotational motions of the associated photospheric bright points.

Although the motions of GBPs are chaotic (Nisenson et al.~\cite{Nisenson03}; M\"{o}stl et al.~\cite{Mostl06}; Chitta et al.~\cite{Chitta12}), it seems there are some typical motion types, such as straight motion, rotary motion and so on. However, studies focusing on motion types of GBPs are relatively scarce. For exploring the characteristics of different motion types, we designed a classification method and categorized motions of GBPs into three types: straight, erratic and rotary motions. The characteristics in terms of horizontal velocity, diffusion, and direction of motion of different motion types were measured and compared. The layout of the paper is as follows. The observations and data reduction are described in Section~\ref{sect:Obs}. The classification method is detailed in Section~\ref{sect:classify}. In Section~\ref{sect:results}, the characteristics of different motion types are presented. Finally, the discussion and conclusion are given in Section~\ref{sect:discussion}.

\section{Observations and data reduction}
\label{sect:Obs}

A series of G-band images were obtained between 18:19 $\rm UT$ and 20:40 $\rm UT$ on 2007 February 19 with the Solar Optical Telescope (SOT; Ichimoto et al.~\cite{Ichimoto04}; Suematsu et al.~\cite{Suematsu08}) on-board the Hinode satellite. The field-of-view (FOV) is 27.7$''\times$27.7$''$ with a pixel size of 0.054$''$. The telescope was pointed at a quiet region at disc center. There are 758 frames over a period of roughly 2\,$\rm h$ and 20\,$\rm min$ with a temporal sample of 11\,$\rm s$. The level-0 data sequence was calibrated and reduced to level-1 using a standard data reduction algorithm fg\_prep.pro distributed in solar software.

Since our intention is to study the motions of GBPs, we designed a high accuracy solar image registration procedure based on a cross-correlation method to correct the satellite drift. It includes seven steps: (1) apply a Tukey window as an apodization function on both reference image $f(x, y)$ and register image $g(x, y)$; (2) calculate the cross-power spectrum that is defined as:$R(u, v)=F(u, v)\ast G(u, v)$, where $F(u, v)$ and $G(u, v)$ are the Fourier transform of $f(x, y)$ and $g(x, y)$ respectively, and $\ast$ donates the complex conjugate operation; (3) obtain the centralized cross-correlation surface $r(x, y)$ by applying the inverse Fourier transform to $R(u, v)$ and centering the zero-frequency component; (4) locate the peak in $r(x, y)$, then measure the distances between the peak and the center of $r(x, y)$, which are the pixel-level displacements of the horizontal and vertical components between $f(x, y)$ and $g(x, y)$; (5) shift $g(x, y)$ over the pixel-level displacements, then repeat steps 1-4 until the displacement is below one pixel; (6) determine the sub-pixel displacements by measuring the centroid of the peak in $r(x, y)$. The modified moment method is applied and the threshold is selected by the minimum value of a small circle surrounding the peak. (7) shift $g(x, y)$ over the sub-pixel displacements by the bicubic interpolation method.

The algorithm is based on the assumption that the maximum of the peak is at the same position as the centroid. It implies the correlation value above the threshold should be symmetrical in the horizontal and vertical directions. This assumption is only acceptable in a very small center area around the cross correlation peak. Our simulated experiment shows that the alignment accuracy of this procedure could be as high as 0.02 pixels between two neighbouring frames if we set the radius of the circle as two pixels. After image alignment, the overlap of the FOV in the sequence is 20.5$''\times$20.5$''$ (380 pixel$\times$380 pixel).

A Laplacian and Morphological Dilation algorithm described in Feng et al. (\cite{Feng12}) was used to detect GBPs in each image, and a three-dimensional (3D) segmentation algorithm described in Yang et al. (\cite{Yang14}) was employed to track the evolution of GBPs in the image sequence.

\begin{figure}
   \centering
   \includegraphics[width=0.75\textwidth, angle=0]{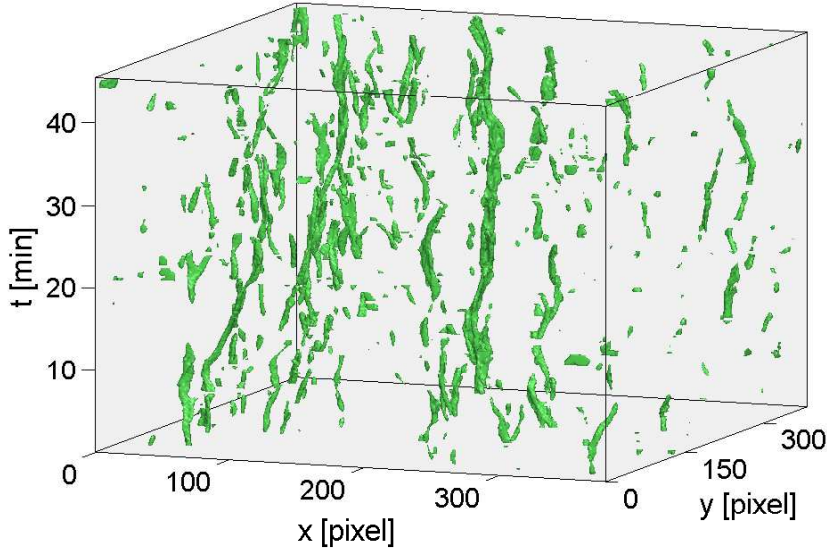}
   \caption{A segment of 3D space-time cube whose size is 380 pixel$\times$380 pixel$\times$46 min. The $x$ and $y$ axes are the two-dimensional image coordinates, and the $t$ axis represents frame index or the time-slice of the image sequence. It can be seen that the GBPs move in different motion types.}
   \label{Fig:fig1}
   \end{figure}

A 3D space-time cube $(x, y, t)$ is modeled where the $\bm{x}$ and $\bm{y}$ axes are the two-dimensional image coordinates, and the $t$ axis represents frame index or the time-slice of the image sequence. The 3D structures in the space-time cube represent the evolution of GBPs. Figure~\ref{Fig:fig1} shows a segment of the 3D space-time cube whose size is 380 pixel$\times$380 pixel$\times$46 min. The chaos scenario is very similar to previous simulations of photospheric turbulent convection and evolution of magnetic footpoints (Carlsson et al.~\cite{Carlsson10}; Moll et al.~\cite{Moll11}; Shelyag et al.~\cite{Shelyag12}; Wedemeyer-B\"{o}hm et al.~\cite{Wedemeyer12}). It can be seen that these GBPs have different motion types during their lifetime. For instance, some GBPs move in nearly straight lines in different directions, and some move in somewhat circular paths. It is worth noting that not only twisted but also braided GBPs can be found in this cube.

\section{Classification of motions}
\label{sect:classify}
Although the motions of GBPs look random, there are some typical types. To explore the characteristics of different motion types, we designed a method to classify the motions.

Initially, some GBPs were discarded for reducing statistical error before classification. A GBP is discarded if (1) its equivalent diameter is smaller than 100\,$\rm km$ or larger than 300\,$\rm km$, (2) its lifetime is shorter than 55\,$\rm s$, (3) its velocity exceeds 7\,$\rm km$ $\rm s^{-1}$, (4) its life cycle is not complete or (5) if merging or splitting occurs during its lifetime. As a result, a total of 753 3D evolving structures remain.

Next, we put stationary GBPs, or those that have limited motion during their lifetimes, aside to measure the dynamics of the GBPs in detail. Projecting the 3D evolving structures onto the two-dimensional $(x, y)$ space, the path of each GBP can be described respectively by the sequence of centroid coordinates $(X(t), Y(t))$ of the GBP in each frame, $t$. Bodn\'{a}rov\'{a} et al. (\cite{Bodn13}) defined a rate of motion as $m =d/r$, where $d$ is the displacement over the lifetime of a GBP, as $d=\sqrt{(X(n)-X(1))^{2}+(Y(n)-Y(1))^{2}}$, here $1$ is the start frame and $n$ is the final frame during its lifetime; $r$ is the radius of the circle which corresponds to the size of the GBP at its start location. Bodn\'{a}rov\'{a} et al. (\cite{Bodn13}) took about 45$\%$ of the tracked GBPs as stationary GBPs with $m <$ 1. However, if a GBP moves in a large circle path, and its start location is very close to its final location, the value $d$ will be very small. This could result in misjudging it as a stationary GBP. Hence, to address this issue, we revised the rate of motion as $m'=d'/r'$, where $d' =\sqrt{(X_{max}-X_{min})^{2}+(Y_{max}-Y_{min})^{2}}$, here $X_{max}$ and $X_{min}$ are the maximum and minimum coordinates of the path of a single GBP in the $\bm{x}$ axis, and $Y_{max}$ and $Y_{min}$ are in the $\bm{y}$ axis; $r'$ is the radius of the circle which corresponds to the maximum size of the GBP during its lifetime. The $d'$ value may represent its actual range of motion more realistically. All $m'$ value of GBPs were calculated and about 43$\%$ of them are below 1, which means their range of motion is smaller than their own maximum radius. About 25$\%$ are greater than 2, which means that these GBPs show significant movement. The maximum value of $m'$ is 7.5. As a result, a total of 324 GBPs with $m' <$ 1 were taken as stationary GBPs. The remaining 429 GBPs with $m'$ $\geq$ 1, called non-stationary GBPs, were involved in classification of motion types.

Finally, we classified the non-stationary GBPs into three motion types. An index of motion type, $mt$, is defined as $mt=d/L$, where $d$ is the displacement defined as above; $L$ is the whole path length, defined as $L=\sum_{t=1}^{n}\sqrt{\triangle X(t)^{2}+\triangle Y(t)^{2}}$, here $\triangle X(t)=X(t+1)-X(t)$, $\triangle Y(t)=Y(t+1)-Y(t)$. The index of motion type is a ratio of the displacement of a GBP to its whole path length.

According to the definition, $mt$ is between 0 and 1. If a GBP moves in a nearly straight line, then $mt$ will be close to 1. If it moves in a nearly closed curve, then $mt$ will be close to 0. In our data set, the maximum $mt$ is 0.996 and the minimum is 0.078.

\begin{figure}
   \centering
   \includegraphics[width=0.9\textwidth, angle=0]{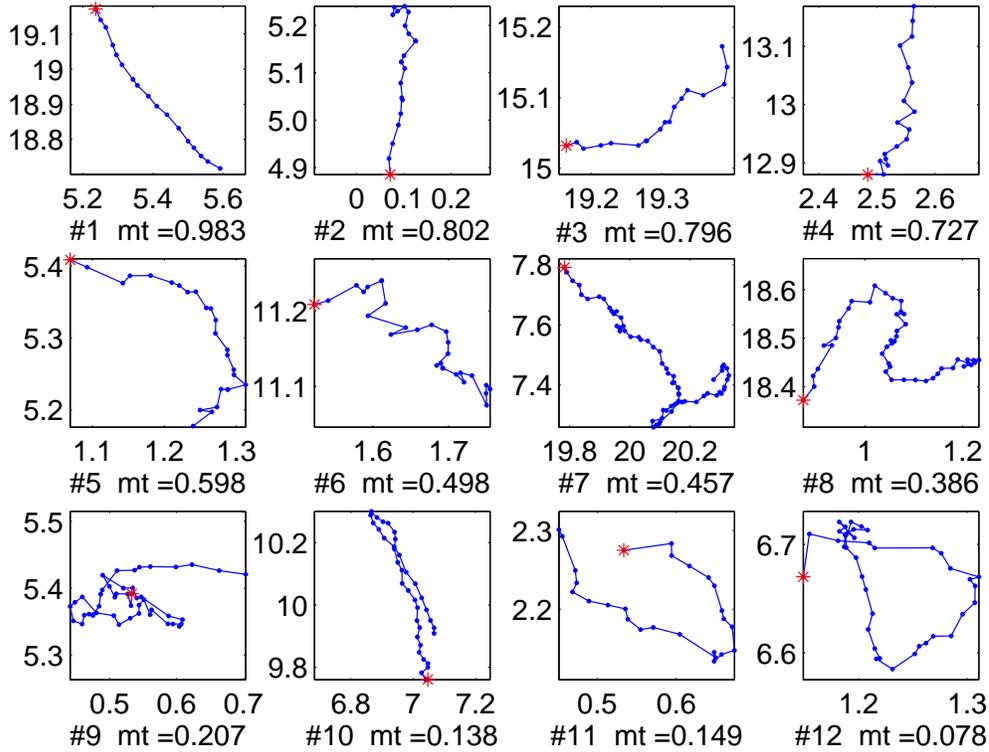}
   \caption{The paths of twelve individual GBPs with corresponding $mt$. The starting position of each path is indicated by a red star . The $x$ and $y$ coordinates are ticked by arcsec.}
   \label{Fig:fig2}
   \end{figure}

Figure~\ref{Fig:fig2} shows the paths of twelve individual GBPs with corresponding $mt$. It can be seen that GBP $\#$ 1 and $\#$ 2 move in nearly straight lines, and their $mt$ values are 0.983 and 0.802 respectively. However, GBP $\#$ 11 and $\#$ 12 rotate in almost closed curve, and their $mt$ values drop to 0.149 and 0.078, respectively. All paths of GBPs are very similar to previous studies (Nisenson et al.~\cite{Nisenson03}; M\"{o}stl et al.~\cite{Mostl06}; Chitta et al.~\cite{Chitta12}). However, we have not found a spiral path similar to those reported by Bonet et al. (\cite{Bonet08}).

\begin{figure}
   \centering
   \includegraphics[width=0.6\textwidth, angle=0]{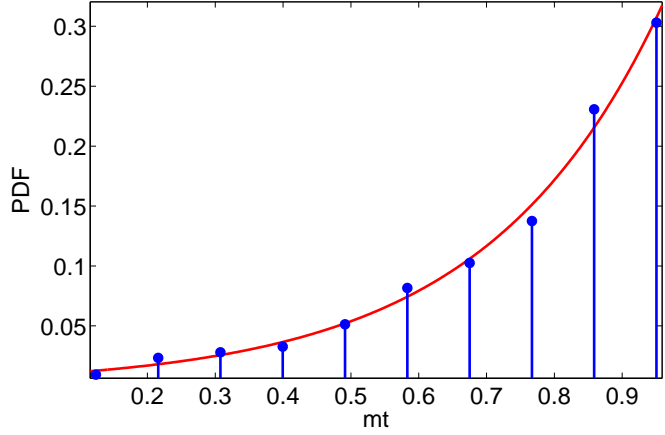}
   \caption{The distribution of $mt$. The PDF of $mt$ is drawn in blue. The red solid line is the exponential fit.}
   \label{Fig:fig3}
   \end{figure}

Figure~\ref{Fig:fig3} shows the Probability Density Function (PDF) of $mt$ and the best exponential fit. The $mt$ value of 50$\%$ of GBPs is larger than 0.83 and 15$\%$ of GBPs have a $mt$ value of less than 0.5. That means most GBPs move in straight or nearly straight lines, and only a few GBPs move in rotary paths. Focusing on straight motion and rotary motion, we classified the motions into three types: straight, erratic and rotary motion types. A GBP which moves in a straight or nearly straight line belongs to the straight motion type, and a GBP which draws a closed or almost closed curve belongs to the rotary motion type. If the path of a GBP is neither straight nor rotary, it will be categorized into erratic motions type. The erratic motion type is more like a transition between straight and rotary type. Considering the exponential distribution of $mt$, and checking individual GBPs, we set one threshold between straight and erratic motion of $mt$ at a value of 0.8 and another between erratic and rotary motion at a value of 0.4. The threshold of 0.8 is reliable for distinguishing the straight motion type. However, the threshold of 0.4 is a compromise. The lower the threshold of $mt$ is, the clearer the rotary motion of the GBP is. Unfortunately, the quantity of GBPs with low $mt$ is very small. If we set a very low threshold, the statistic result would be meaningless. Consequently, 54$\%$ of GBPs  (232 GBPs) were categorized as having straight motions, about 7$\%$ (31 GBPs) had rotary motions, and the others were erratic in nature. Comparing the paths in different panels in Figure~\ref{Fig:fig2}, we could see that the classification is consistent with our subjective understanding. Some special cases, will be discussed in Section ~\ref{sect:discussion}.

\section{Results}
\label{sect:results}

\subsection{Horizontal Velocity}
\label{sect:velocity}
\begin{figure}
   \centering
   \includegraphics[width=0.6\textwidth, angle=0]{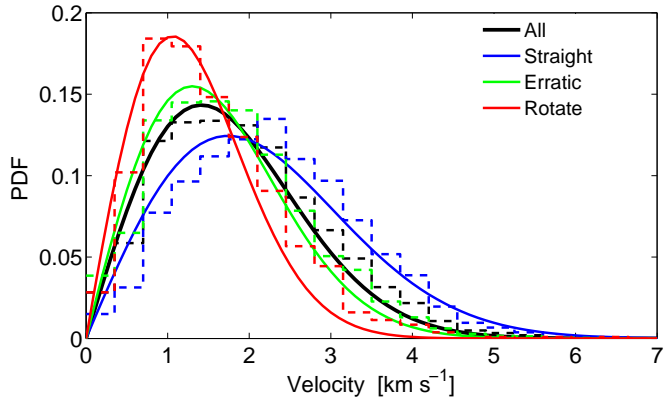}
   \caption{{\small The PDFs of straight, erratic and rotary motion type are shown in blue, green and red dotted lines, respectively. The PDFs are curve fitted with a Rayleigh curve which is drawn with solid lines in corresponding color, respectively. The mathematical expectations of velocities and standard errors are 2.18$\pm$0.08, 1.63$\pm$0.09 and 1.33$\pm$0.07\,$\rm km$ $\rm s^{-1}$, respectively. Besides that, the PDF and curve fit for all motion types as a whole is drawn in black, with mathematical expectation of 1.77$\pm$0.08\,$\rm km$ $\rm s^{-1}$.}}
   \label{Fig:fig4}
   \end{figure}

The horizontal velocity of a GBP between two neighbouring frames can be calculated by $v=\sqrt{v_{x}^{2}+v_{y}^{2}}$. The horizontal velocities of all the GBPs studied were calculated and the corresponding PDFs all follow Rayleigh distributions. Figure~\ref{Fig:fig4} shows the PDFs for the horizontal velocities of the different motion types in dotted lines and the corresponding curve fits for these are displayed as solid lines. The mathematical expectations of velocities and standard errors are 2.18$\pm$0.08\,$\rm km$ $\rm s^{-1}$ for straight motion type, 1.63$\pm$0.09\,$\rm km$ $\rm s^{-1}$ for erratic motion type and 1.33$\pm$0.07\,$\rm km$ $\rm s^{-1}$ for rotary motion type, respectively. For comparison, the PDF and Rayleigh fit for all motion types as a whole are drawn in black, with mathematical expectation of 1.77$\pm$0.08\,$\rm km$ $\rm s^{-1}$.

To quantify the difference between different motion types, we performed a two sample Kolmogorov-Smirnov (K-S) test between the distributions of straight and rotary motion types at 5$\%$ significance level. The test result indicates that the difference of horizontal velocities between them is significant.

\begin{figure}
   \centering
   \includegraphics[width=0.6\textwidth, angle=0]{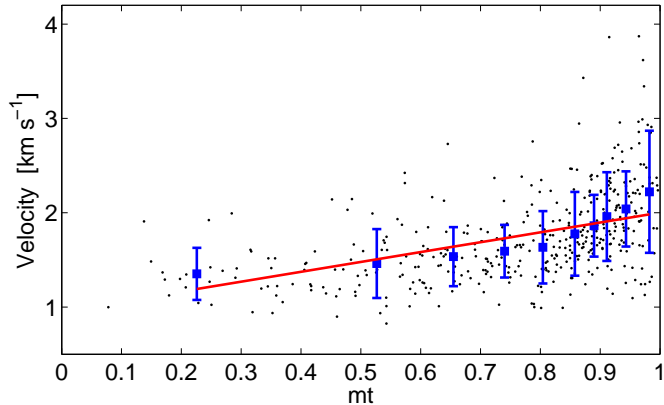}
   \caption{{\small The scatter diagram associated with the indices of motion type of 429 GBPs versus their mean horizontal velocities are drawn in black. All GBPs were divided into 10 equal bins. The mean values and associated standard deviations of velocity belonging to each bin are drawn in blue. The red solid line is the best linear fit for the means.}}
   \label{Fig:fig5}
   \end{figure}

We also analyzed the correlation between the indices of motion type and horizontal velocities. The mean horizontal velocity of each GBP during its lifetime was calculated. Figure~\ref{Fig:fig5} shows the scatter diagram associated with the indices of motion type of 429 GBPs versus their mean horizontal velocities in black. The mean values and the associated standard deviation of the horizontal velocities were calculated separately and displayed in blue after all GBPs were divided into 10 equal bins. The best linear fit for the mean values is displayed as a red solid line. Additionally, we calculated the correlation coefficient between $mt$ and the mean horizontal velocity, and obtained a value of 0.88. The results of classification and correlation analysis correspond to one another, and both of them suggest that a GBP moving in a straight line has a higher velocity, and vice versa.

Furthermore, we quantified the evolution of GBPs for the different motion types in terms of velocity using a method described in Yang et al. (\cite{Yang14}). The basis of the method is lifetime normalization which standardizes different lifetimes to common stages. In general, dividing the life into five stages is representative of the entire evolution process, which can be described as birth, growth, middle, decay and disappearance. Significantly, the authors illuminated that the GBPs have a similar evolutional pattern in terms of velocity, no matter how many stages that the GBPs are divided into during their lifetime. Therefore, we divided the life of each GBP into five stages for different motion types respectively. The PDFs of velocities in five stages are each separately fitted to a Rayleigh distribution. The mathematical expectations with associated standard errors are shown in Table~\ref{Tab:tab1}. For comparison, the result of all motion types as a whole are listed. It can be seen that a GBP moving in a straight line not only has a higher velocity, but also has an almost constant velocity (about 2\,$\rm km$ $\rm s^{-1}$). However, a GBP moving in a rotary path has a lower and variational velocity. On average, it moves relatively fast (1.36\,$\rm km$ $\rm s^{-1}$) at birth, then decreases to its lowest velocity (1.20\,$\rm km$ $\rm s^{-1}$) in the middle stage, and accelerates again (1.58\,$\rm km$ $\rm s^{-1}$) in the decay stage until it disappears. Interestingly, Balmaceda et al. (\cite{Balmaceda10}) described a similar phenomenon in analyzing a series of images at different atmospheric layers. They reported the stochastic evolution of granules that allow the bright points to approach the vortex influence, increase their velocities, and eventually fall into the vortex.

\begin{table}
\begin{center}
\caption[]{The Mathematical Expectations and Standard Errors of the Velocity of GBPs Belonging to Different Motion Types after Dividing Their Lives into Five Stages}\label{Tab:tab1}
  \begin{tabular}{lccccc}
  \hline\noalign{\smallskip}
  Motion Type & Birth stage & Growth stage & Middle stage & Decay stage & Disappearance stage \\
  &[$\rm km$ $\rm s^{-1}$]    &[$\rm km$ $\rm s^{-1}$]  &[$\rm km$ $\rm s^{-1}$]  &[$\rm km$ $\rm s^{-1}$]  &[$\rm km$ $\rm s^{-1}$]\\
  \hline\noalign{\smallskip}
Straight   &2.17$\pm$0.13   &2.16$\pm$0.13  &2.16$\pm$0.12 &2.17$\pm$0.14  &2.18$\pm$0.13  \\
Erratic    &1.64$\pm$0.12	&1.56$\pm$0.11  &1.51$\pm$0.11 &1.54$\pm$0.11  &1.76$\pm$0.13  \\
Rotary     &1.36$\pm$0.18	&1.21$\pm$0.16	&1.20$\pm$0.16 &1.29$\pm$0.17  &1.58$\pm$0.21  \\
All        &1.83$\pm$0.08	&1.72$\pm$0.08	&1.61$\pm$0.08 &1.77$\pm$0.08  &1.86$\pm$0.09  \\

  \noalign{\smallskip}\hline
\end{tabular}
\end{center}
\end{table}

\subsection{Diffusion}
\label{sect:diffusion}

\begin{figure}
   \centering
   \includegraphics[width=0.6\textwidth, angle=0]{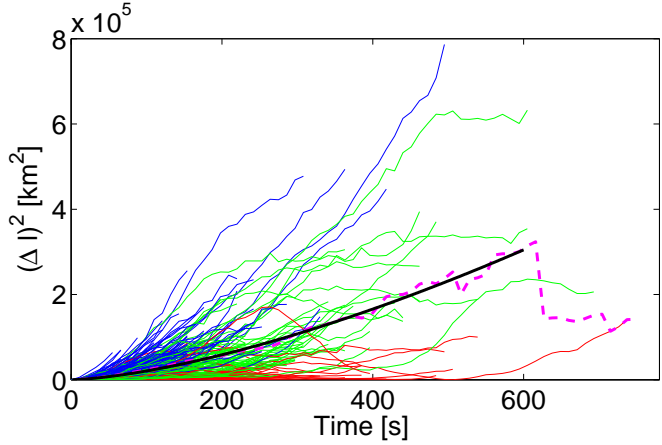}
  \caption{{\small The mean-squared displacement of straight, erratic and rotary motion type are shown in blue, green and red lines, respectively. The mean-squared displacement for all motion types as a whole, with time, is displayed as a dashed pink line. Black lines show the best power fit with $\gamma$ = 1.50 for time scales $\tau <$ 600\,$\rm s$, which is super-diffusive.}}
   \label{Fig:fig6}
   \end{figure}

Photospheric diffusion processes represent the efficiency of GBP dispersal in the photosphere, which can be characterised by the relation $\langle(\Delta l)^{2}\rangle = C\tau^{\gamma}$, where $\langle(\Delta l)^{2}\rangle$ represents the mean-squared displacement of the tracked GBP between its location at any given time $\tau$ and its initial position; $\gamma$ is the diffusion index and $C$ is a constant of proportionality. Motions with $\gamma < 1$, $\gamma = 1$ and $\gamma > 1$ are called sub-diffusive, normal-diffusive (random walk) and super-diffusive, respectively. The squared displacement $(\Delta l)^{2}$ of the three motion types were calculated separately and illustrated in Figure~\ref{Fig:fig6} in different colors. Statistically, the behaviour of all motion types as a whole is characterized by the scaling of the mean-squared displacement covered with time (dashed pink line). Only three GBPs can be traced beyond 600\,$\rm s$, so the extended tail of the mean-squared displacement is truncated to aid in the depiction of the diffusion index. Black line shows the best power fit with $\gamma$ = 1.50$\pm$0.08 inside the time interval of 11\,$-$\, 600\,$\rm s$, which displays significant super-diffusive behavior. Interestingly, most GBPs which move in straight lines have squared displacement above the fit line. These GBPs drift fast and have more L\'{e}vy flights, and they make a primary contribution to super, or even ballistic diffusion ($\gamma =$ 2). However, most GBPs which move in rotary paths have squared displacement below the fit line. They move slow and contribute to a sub-diffusive regime.

\begin{figure}
   \centering
   \includegraphics[width=0.6\textwidth, angle=0]{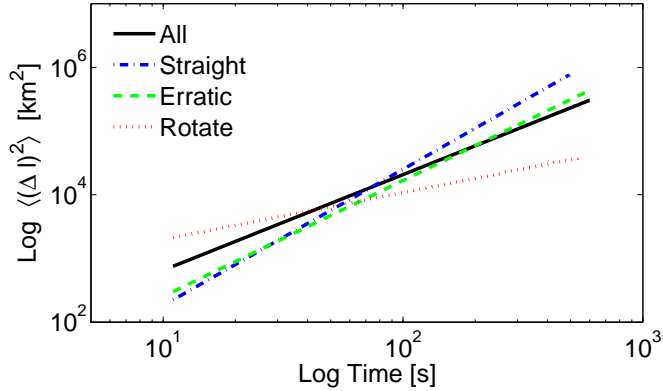}
  \caption{{\small Mean-squared displacement $\langle(\Delta l)^{2}\rangle$ as a function of time, $\tau$, on a log-log scale. The best linear fits for straight, erratic and rotary motions are displayed in blue (dash-dotted line), green (dashed line) and red (dotted line), with $\gamma$ = 2.13$\pm$0.09, 1.82$\pm$0.07 and 0.73$\pm$0.19, respectively. These are fitted for times $\tau <$ 495\,$\rm s$, $\tau<$ 600\,$\rm s$ and $\tau<$ 550\,$\rm s$, respectively. A solid line shows the best fit for all motion types as a whole in black, with $\gamma$ = 1.50$\pm$0.08 for $\tau <$ 600\,$\rm s$.}}
   \label{Fig:fig7}
   \end{figure}

We also calculated the mean-squared displacement for GBPs belonging to straight, erratic and rotary motion type respectively. We found values for $\gamma$ of 2.13$\pm$0.09, 1.82$\pm$0.07 and 0.73$\pm$0.19 determined for time scales $\tau <$ 495\,$\rm s$, $\tau <$ 600\,$\rm s$ and $\tau <$ 550\,$\rm s$, respectively. The mean-squared displacement $\langle(\Delta l)^{2}\rangle$ as a function of time, $\tau$, of the three motion types are displayed in Figure~\ref{Fig:fig7} on a log-log scale. For comparison, mean-squared displacement of all motion types as a whole is also drawn in Figure~\ref{Fig:fig7} with $\gamma $= 1.50.

The diffusion coefficient, $K$, representing the rate of area in unit time that a GBP moves across, is estimated as a function of timescale by Monin et al. (\cite{Monin75}), $K(\tau) = \frac{C\gamma}{4}\tau^{\gamma-1}$. In practice, $C$ is the constant term of the equation $\langle(\Delta l)^{2}\rangle = C\tau^{\gamma}$, $\gamma$ is diffusion index and $\tau$ represents the lifetime of the GBP. As a result, $K$ is 850$\pm$37\,$\rm km^{2}$ $\rm s^{-1}$ for straight motion type, 331$\pm$24\,$\rm km^{2}$ $\rm s^{-1}$ for erratic motion type and 13$\pm$9\,$\rm km^{2}$ $\rm s^{-1}$ for rotary motion type, respectively. Considering all motion types as a whole, $K$ is 191$\pm$20\,$\rm km^{2}$ $\rm s^{-1}$. The results for all three motion types are consistent with the results found for velocity and $\gamma$.

\subsection{Direction of Motion}
\label{sect:direction}

\begin{figure}
   \centering
   \includegraphics[width=0.6\textwidth, angle=0]{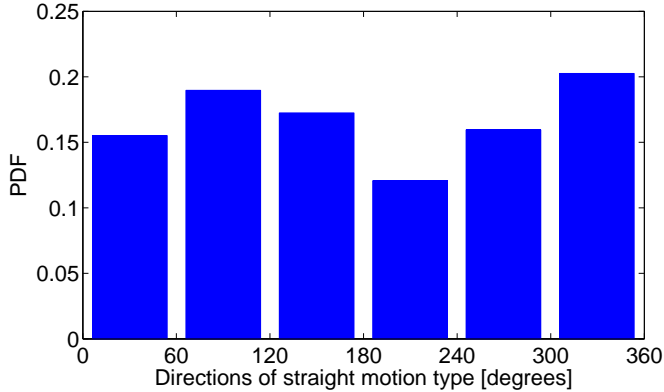}
   \caption{The distribution of the start-to-end direction angle of GBPs for the straight motion type grouped in six bins. The $x$ axis is ticked in degrees. The directions of motion are homogeneous and isotropical.}
   \label{Fig:fig8}
   \end{figure}

In order to explore whether GBPs drift in some preferential directions, a start-to-end direction angle is defined as the angle made by a given line (connecting sub-pixel centroids of the brightness of a GBP at its start location and its final location) with the reference axis. We calculated start-to-end direction angle of each GBP belonging to straight motion type. Figure~\ref{Fig:fig8} is the distribution of direction angles grouped in six bins. Obviously, the distribution does not show GBPs drifting in any preferential direction. Instead, their directions of motion are homogeneous and isotropical.

\begin{figure}
   \centering
   \includegraphics[width=0.6\textwidth, angle=0]{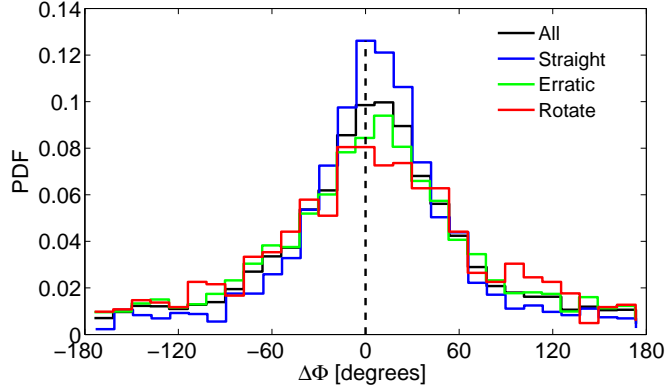}
   \caption{The PDFs of $\bigtriangleup\emptyset$ for straight, erratic and rotary motion type in blue, green and red, respectively. The black line shows the PDF for all motion types as a whole. The $x$ axis is ticked in degrees.}
   \label{Fig:fig9}
   \end{figure}

In relation to previous studies (Nisenson et al.~\cite{Nisenson03}; Bodn\'{a}rov\'{a} et al.~\cite{Bodn13}), we also measured the change of direction angle, $\bigtriangleup\emptyset$, between neighbouring frames. The PDFs of $\bigtriangleup\emptyset$ of the different motion types are drawn in Figure~\ref{Fig:fig9}. The fact that all distributions peak at 0 indicates that the direction of motion between neighbouring frames varies very slowly. Since it is calculated by sub-pixel centroid, the direction of motion will only gradually change. However, the distributions of different motion types are not exactly the same. We employed kurtosis to measure the peakedness of the PDFs in Figure~\ref{Fig:fig9}. Kurtosis is defined as following:

\begin{equation}
        kurtosis=\frac{\frac{1}{n}\sum_{i=1}^{n}(x_{i}-\overline{x})^{4}}{(\frac{1}{n}\sum_{i=1}^{n}(x_{i}-\overline{x})^{2})^{2}}-3.
\end{equation}

  The kurtosises of straight, erratic and rotary motion type are 1.25, 0.17, -0.11, respectively. The distribution of straight motion type is significantly leptokurtic (kurtosis$>$0), which has a more acute peak around the mean with thinner tails, and actually 54$\%$ of $\bigtriangleup\emptyset$ are limited between -30 and 30 degrees. However, the distribution of rotary type is platykurtic (kurtosis$<$0), which has a lower and wider peak around the mean with larger tails. Additionally, the kurtosis of all motion types as a whole is 0.42, which is slightly leptokurtic.

\section{Discussion and conclusion}
\label{sect:discussion}

We have analyzed the motions of isolated GBPs in the quiet Sun using a high resolution G-band image sequence acquired with Hinode/SOT. Although the motions of GBPs look random, they can be classified into different types. For exploring the motion of GBPs in detail, a rate of motion defined by Bodn\'{a}rov\'{a} et al. (\cite{Bodn13}) is revised as a ratio ($m'$) of the real range of motion to the maximum radius of the GBP if we define it as having circular geometry. About 43$\%$ of GBPs studied displaying $m' <$ 1 are put aside as stationary GBPs, which means that these GBPs do not move beyond their own boundaries during their lifetimes. The remaining 429 non-stationary GBPs are classified based on an index of motion type, $mt$, which is defined as a ratio of the displacement to the whole path length. The exponential distribution of these indices shows that most GBPs follow straight or nearly straight lines, with only a few moving in rotary paths. Consequently, we categorize the non-stationary GBPs into three types by imposing two subjective thresholds. About 54$\%$ of GBPs are categorized into a straight motion type, about 7$\%$ into a rotary motion, and the others fall into a motion type we define as erratic.

Our classification method is simple but effective. It allows us to distinguish straight and rotary motions and allows us to explore the characteristics of different motion types. However, the motions of GBPs are very complicated, so the boundaries between these types are not quite clear. For example, GBP $\#$ 7 in Fig~\ref{Fig:fig2} moves in a straight line at first then has a circular path at the end of its lifetime. Considering its $mt$ value and the proportion of the rotary motion in its whole path of motion, the erratic motion type is a suitable choice. Both GBP $\#$ 5 and $\#$ 11 move in opened curve paths, but they are categorized into different types based on their displacement. In fact, the expansion and evolution of granules may result in the interruption of a rotary path of a GBP, so classifying a GBP with an open but clear curve into the rotary motion type might be reasonable. Additionally, if a GBP changes its direction frequently as $\#$ 9, or moves from side to side with a small displacement, such as $\#$10, these will be categorized into the rotary motion type too.

The mean velocity of all motion types is 1.77\,$\rm km$ $\rm s^{-1}$. Earlier studies of velocity are listed in Table~\ref{Tab:tab2}. Our result is slightly elevated in comparison to previously published studies. One reason may be attributed to the data, with different spatial and temporal resolution/sampling across the various studies, and the fact that GBPs are identified with different methods in each of the studies (Utz et al.~\cite{Utz10}). Utz et al. (\cite{Utz10}) reported 1-2\,$\rm km$ $\rm s^{-1}$ for different spatial and temporal sampling, and indicated that higher spatial and temporal resolution results in a higher velocity. Another reason may be that some GBPs described in Section~\ref{sect:classify} and stationary GBPs with $m' <$ 1 are not counted. If all isolated GBPs are counted, the mean velocity is 1.66$\pm$0.07\,$\rm km$ $\rm s^{-1}$.

\begin{table}
\begin{center}

\caption[]\centering{Overview of Mean Velocities of GBPs in Previous Studies}\label{Tab:tab2}
  \begin{tabular}{lccc}
  \hline\noalign{\smallskip}
  Paper & Pixel size & Temporal sample & Mean velocity \\
  \hline\noalign{\smallskip}
Muller et al. (\cite{Muller94})         &  0.25$''$     & 20\,$\rm s$    &   1.33\,$\rm km$ $\rm s^{-1}$ \\
Berger et al. (\cite{Berger98b})        & 0.4$''$       & 23.75\,$\rm s$ &   1.47\,$\rm km$ $\rm s^{-1}$   \\
Nisenson et al. (\cite{Nisenson03})     & 0.071$''$     & 30\,$\rm s$    &   1.31\,$\rm km$ $\rm s^{-1}$  \\
M\"{o}stl et al. (\cite{Mostl06})       & 0.041$''$     & 20\,$\rm s$    &   1.11\,$\rm km$ $\rm s^{-1}$  \\
Keys et al. (\cite{Keys11})             & 0.069$''$     & 2\,$\rm s$     &   $\sim$1\,$\rm km$ $\rm s^{-1}$  \\
Utz et al. (\cite{Utz10})               & 0.054$''$     & 11\,$\rm s$     &  1-2\,$\rm km$ $\rm s^{-1}$  \\
Bodn\'{a}rov\'{a} et al. (\cite{Bodn13})& 0.071$''$     & 30\,$\rm s$    &   0.89\,$\rm km$ $\rm s^{-1}$  \\
This study                              & 0.054$''$     & 11\,$\rm s$    &   1.77\,$\rm km$ $\rm s^{-1}$  \\

  \noalign{\smallskip}\hline
\end{tabular}
\end{center}
\end{table}

The velocity of GBPs for the different motion types are separately calculated and compared in this study. The mathematical expectations and standard errors are 2.18$\pm$0.08\,$\rm km$ $\rm s^{-1}$ for straight motion type, 1.63$\pm$0.09\,$\rm km$ $\rm s^{-1}$ for erratic motion type and 1.33$\pm$0.07\,$\rm km$ $\rm s^{-1}$ for rotary motion type, respectively. The horizontal velocity has a notable positive correlation with the index of motion type (correlation coefficient is 0.88). Both classification and correlation analysis suggest that if a GBP moves in a straight line, it will move fast; however, if a GBP moves in a rotary path, it will move slowly. We also demonstrate that there is a significant difference between straight and rotary motion types by a two-sample K-S test. Importantly, GBPs with higher velocity ($>$ 2\,$\rm km$ $\rm s^{-1}$) have been linked to the production of magneto-sonic kink waves (de Wijn et al.~\cite{de09}). As shown previously, these kink waves may act as a conduit for imparting energy into the upper solar atmosphere. About 28$\%$ of GBPs studied exceed 2\,$\rm km$ $\rm s^{-1}$. M\"{o}stl et al. (\cite{Mostl06}) measured 11.3$\%$ and Keys et al. (\cite{Keys14}) found about 15$\%$ have velocities greater than 2\,$\rm km$ $\rm s^{-1}$. Our higher proportion is probably the result of the higher velocity stated above.

The diffusion index, $\gamma$, is 1.50$\pm$0.08 averaged over all non-stationary, isolated GBPs. Recent authors have agreed on a super-diffusive dynamic regime. Abramenko et al. (\cite{Abramenko11}) found $\gamma$ = 1.53 with high resolution TiO observations of a quiet Sun area. Chitta et al. (\cite{Chitta12}) reported $\gamma$ of 1.59 for their short-lived GBPs. Jafarzadeh et al. (\cite{Jafarzadeh14}) calculated the squared-displacement of each GBP and derived $\gamma$ = 1.69$\pm$0.08 by the distribution of the diffusion indices. Giannattasio et al. (\cite{Giannattasio14}) found $\gamma$ = 1.27$\pm$0.05 and $\gamma$ = 1.08$\pm$0.11 in network (at smaller and larger scales, respectively), and $\gamma$ = 1.44$\pm$0.08 in internetwork regions. Keys et al. (\cite{Keys14}) estimated $\gamma$ of $\sim$1.2 for both quiet region and active region, then confirmed that the dynamic properties of GBPs arise predominately from convective motions. Our result is in qualitative agreement with recent studies of diffusion. However, earlier studies, almost all other authors interpreted their results as indicative of normal- or sub-diffusive processes. Berger et al. (\cite{Berger98a}) found indications of slight super-diffusivity among otherwise normal diffusive GBPs in network regions. Cadavid et al. (\cite{Cadavid99}) found that although the motion of magnetic network GBPs in the photosphere is random if their lifetimes are larger than 25 minutes, GBPs with lifetimes less than 20 minutes migrate sub-diffusively. All authors suggested that estimations of the diffusion index may be effected by the temporal and spatial scales.

We find that $\gamma$ is 2.13$\pm$0.09 for straight motion type, 1.82$\pm$0.07 for erratic motion type, and 0.73$\pm$0.19 for rotary motion type, respectively. The larger value of $\gamma$ for straight motion type come from the fact that these GBPs move continuously in the same direction and have more L\'{e}vy flights in their path. On average, they move faster and accelerate with time. The straight motion makes a large contribution to super, or even ballistic diffusion. However, the smaller the $\gamma$ value of rotary motion type may be due to the fact that it changes its general direction and moves closer to its initial coordinates again. This results in a sub-diffusive regime. The $\gamma$ values of different motion types confirm that our classification method is reasonable. Similarly, Jafarzadeh et al. (\cite{Jafarzadeh14}) proposed mean-squared displacement can result in the mixing of different diffusive processes, and therefore calculated the trajectory of each BP and described it by a diffusion index. They categorized GBPs into different motion type based on diffusion index.

The coefficient $K$ indicates the efficiency of the dispersion. The $K$ values are 850$\pm$37\,$\rm km^{2}$ $\rm s^{-1}$ for straight motion type, 331$\pm$24\,$\rm km^{2}$ $\rm s^{-1}$ for erratic motion type, and 13$\pm$9\,$\rm km^{2}$ $\rm s^{-1}$ for rotary motion type, respectively. We also measure $K$ of all motion types as a whole of 191$\pm$20\,$\rm km^{2}$ $\rm s^{-1}$. GBPs which move in straight lines experience the quickest rate of diffusion, while GBPs which move in rotary paths experience the slowest rate of diffusion. Our result confirms and furthers previous studies. Berger et al. (\cite{Berger98a}) determined $K$ = 60.4$\pm$10.9\,$\rm km^{2}$ $\rm s^{-1}$ for network MBPs by assuming normal-diffusion (i.e. $K = sd/2d\tau$), whereas the $\gamma$ value of 1.34$\pm$0.06 infers a super-diffusive regime. Utz et al. (\cite{Utz10}) found $K$ = 350$\pm$20$\,\rm km^{2}$ $\rm s^{-1}$. Manso Sainz et al. (\cite{Manso11}) measured $K$ = 195\,$\rm km^{2}$ $\rm s^{-1}$ for footpoints of small-scale internetwork magnetic loops. Jafarzadeh et al. (\cite{Jafarzadeh14}) found $K$ = 257$\pm$32\,$\rm km^{2}$ $\rm s^{-1}$ averaged over all MBPs. Giannattasio et al. (\cite{Giannattasio14}) estimated that the diffusivity increases from $\sim$ 100 to $\sim$ 400\,$\rm km^{2}$ $\rm s^{-1}$ along the time scales (100\,$\rm s$ $< \tau <$ 10000\,$\rm s$). The sources of such a large range could be the result of different features (magnetic elements, bright points), different regions (quiet region, active region), and/or different temporal and spatial scales etc.

The distribution of the directions of motion of GBPs that move in straight lines show no particular preferential flow direction. It is homogeneous and isotropical.  Verma et al. (\cite{verma11}) adapted LCT to measure horizontal flow fields of GBPs and other features. Keys et al. (\cite{Keys14}) tracked the motions of GBPs for exploring preferential flow directions. Our result is in agreement with them. As well as this result, we also confirm the conclusion of Nisenson et al. (\cite{Nisenson03}) and Bodn\'{a}rov\'{a} et al. (\cite{Bodn13}) in that the direction of motion usually persists between neighbouring frames. Both studies utilized G-band images taken with the Dutch Open Telescope with a temporal sampling of 30\,$\rm s$. We strengthen their claim using the G-band images acquired from the Hinode/SOT with temporal sampling of 11\,$\rm s$. Additionally, we have discussed that the kurtosises of the distributions in the three motion types are different. It is leptokurtic for straight motion type and platykurtic for rotary motion type. That means a GBP which moves in a straight line is more likely to keep its direction than a GBP which moves in a rotary path.

\begin{acknowledgements}
The authors are grateful to the anonymous referee for constructive comments and detailed suggestions to this manuscript. The authors are grateful to the support received from the National Natural Science Foundation of China (No: 11303011, 11263004, 11163004, U1231205), Open Research Program of the Key Laboratory of Solar Activity of the Chinese Academy of Sciences (No: KLSA201414, KLSA201309). This work is also supported by the Opening Project of Key Laboratory of Astronomical Optics \& Technology, Nanjing Institute of Astronomical Optics \& Technology, Chinese Academy of Sciences (No. CAS-KLAOT-KF201306) and the open fund of the Key Laboratory of Modern Astronomy and Astrophysics, Nanjing University, Ministry of Education, China. The authors are grateful to the Hinode team for the possibility to use their data. Hinode is a Japanese mission developed and launched by ISAS/JAXA, collaborating with NAOJ as a domestic partner, NASA and STFC (UK) as international partners. Scientific operation of the Hinode mission is conducted by the Hinode science team organized at ISAS/JAXA. This team mainly consists of scientists from institutes in the partner countries. Support for the post launch operation is provided by JAXA and NAOJ (Japan), STFC (U.K.), NASA (U.S.A.), ESA, and NSC (Norway).
\end{acknowledgements}

\label{lastpage}


\begin{thebibliography}{99}

  \bibitem[2011]{Abramenko11} Abramenko, V. I., Carbone, V., Yurchyshyn, V., et al., 2011, \apj, 743, 133

  \bibitem[2010]{Balmaceda10} Balmaceda, L., Vargas Dom\'{i}nguez, S., Palacios, J., Cabello, I., Domingo, V., 2010, \aap, 513, 6

  \bibitem[1998a]{Berger98a} Berger, T. E., L\"{o}fdahl M. G., Shine R. S., Title  A. M., 1998a, \apj, 506, 439

  \bibitem[1998b]{Berger98b} Berger, T. E., L\"{o}fdahl, M. G., Shine, R. A., Title, A. M., 1998b, \apj, 495, 973

  \bibitem[2013]{Bodn13} Bodn\'{a}rov\'{a}, M., Utz, D., Ryb\'{a}k, J., 2013, \solphys, 289, 5

  \bibitem[2008]{Bonet08} Bonet, J., M\'{a}rquez, I., S\'{a}nchez Almeida, J., Cabello, I., Domingo, V., 2008, \solphys, 687, 131

  \bibitem[1999]{Cadavid99} Cadavid, A. C., Lawrence, J. K., Ruzmaikin, A. A., 1999, \apj, 521, 844

  \bibitem[2010]{Carlsson10} Carlsson, M., Hansteen, V. H., Gudiksen, B. V., 2010, Mem. Soc. Astron. Italiana, 81, 582

  \bibitem[2012]{Chitta12}Chitta, L. P., van Ballegooijen, A. A., Rouppe van der Voort, L., DeLuca, E. E., Kariyappa, R., 2012, \apj, 752, 48

  \bibitem[2002]{Cranmer02} Cranmer S. R., 2002, Space Sci. Rev., 101, 229

  \bibitem[2005]{Cranmer05} Cranmer S. R., van Ballegooijen A. A., 2005, \apjs, 156, 265

  \bibitem[2009]{de09} de Wijn A. G., Stenflo J. O., Solanki S. K., Tsuenta S., 2009, Space Sci. Rev., 144, 275

  \bibitem[2011]{Fedun11} Fedun, V., Shelyag, S., Erd\'{e}lyi, R. 2011, \apj, 727, 17

  \bibitem[2012]{Feng12} Feng S., Ji K. F., Deng H., Wang F., Fu X. D., 2012, Journal of the Korean Astronomical Society, 45, 167

  \bibitem[2013]{Giannattasio13} Giannattasio, F., Del Moro, D., Berrilli, F., et al., 2013, \apj, 770, L36

  \bibitem[2014]{Giannattasio14} Giannattasio, F., Stangalini, M., Berrilli, F., Del Moro, D., Bellot Rubio, L., 2014, \apj, 788, 137

   \bibitem[2010]{Goode10}  Goode, P. R., Yurchyshyn, V., Cao, W., Abramenko, V., Andic, A., Ahn, K., Chae, J. 2010, \apjl, 714, 1, L31

  \bibitem[2004]{Ichimoto04} Ichimoto K., Tsuneta S., Suematsu Y., et al., 2004, in Optical, Infrared, and Millimeter Space Telescopes, ed. J. C. Mather, Proc. SPIE, 5487, 1142

  \bibitem[2014]{Jafarzadeh14} Jafarzadeh, S., Cameron, R. H., Solanki, S. K., et al., 2014, \aap, 563, 101

 \bibitem[2011]{Keys11} Keys P. H., Mathioudakis M., Jess D. B., Shelyag S., Crockett P. J., Christian D. J., Keenan F. P., 2011, \apjl, 740, 40

  \bibitem[2014]{Keys14} Keys, P. H., Mathioudakis, M., Jess, D. B., Mackay, D. H., Keenan, F. P., 2014, \aap, 566, 99

 \bibitem[2006]{Klimchuk06} Klimchuk J. A., 2006, \solphys, 234, 41

 \bibitem[2011]{Manso11} Manso Sainz, R., Mart\'{\i}nez Gonz\'{a}lez, M. J., Asensio Ramos, A., 2011, \aap, 531, 9

 \bibitem[2011]{Moll11} Moll, R., Cameron, R. H., Sch\"{u}ssler, M., 2011, \aap, 533, 126

 \bibitem[1975]{Monin75} Monin, A. S., Iaglom, A. M., 1975, Statistical fluid mechanics: Mechanics of turbulence (Cambridge, MA: MIT Press), 2

 \bibitem[2006]{Mostl06} M\"{o}stl C., Hanslmeier A., 2006, \solphys, 237, 13

  \bibitem[1994]{Muller94} Muller R., Roudier Th., Vigneau J., Auffret H., 1994, \aap, 283, 232

 \bibitem[2003]{Nisenson03} Nisenson, P., van Ballegooijen A. A., de Wijn, A. G., S\"{u}tterlin, P., 2003, \apj, 587, 458

  \bibitem[2011]{Shelyag11} Shelyag, S., Keys, P. H., Mathioudakis, M., Keenan, F. P., 2011, \aap, 526, 5

  \bibitem[2012]{Shelyag12} Shelyag, S., Mathioudakis, M., Keenan, F. P., 2012, \apj, 753, 22

  \bibitem[2013]{Shelyag13} Shelyag, S., Cally, P. S., Reid, A., Mathioudakis, M., 2013, \apjl, 776, L4

  \bibitem[2008]{Suematsu08} Suematsu Y., Tsuneta S., Ichimoto K., et al., 2008, \solphys, 249, 197

  \bibitem[1998]{van98} van Ballegooijen, A. A., Nisenson, P., Noyes, R. W., L\"{o}fdahl, M. G., Stein, R. F., Nordlund, {\AA}
., Krishnakumar, V. 1998, \apj, 509, 435

 \bibitem[2011]{verma11} Verma, M., Denker, C., 2011, \aap, 529, 153

  \bibitem[2005]{Vogler05} V\"{o}gler, A., Shelyag, S., Sch\"{u}ssler, M., Cattaneo, F., Emonet, T., Linde, T. 2005, \aap, 429, 335

  \bibitem[2010]{Utz10} Utz D., Hanslmeier A., Muller R., Veronig A., Ryb\'{a}k J., Muthsam H., 2010, \aap, 511, 39

  \bibitem[2012]{Wedemeyer12} Wedemeyer-B\"{o}hm, S., Scullion, E., Steiner, O., Rouppe van der Voort, L., de La Cruz Rodriguez, J., Fedun, V., Erd\'{e}lyi, R., 2012, Nature, 486, 505

  \bibitem[2014]{Yang14} Yang Y. F., Lin J. B., Ji K. F., Feng S., Deng H., Wang F., 2014, \raa, 6, 741

  \bibitem[2009]{Zhao09} Zhao M., Wang J. X., Jin C. L., Zhou G. P., 2009, \raa, 8, 933

\end{thebibliography}
\end{document}